\documentclass[oldversion]{aa} 
\usepackage{amsmath}
\usepackage{amssymb}
\usepackage{natbib}
\bibpunct{(}{)}{;}{a}{}{,}
\usepackage{graphicx}
\usepackage{txfonts}
\usepackage[english]{babel}
\usepackage{hyperref}
\usepackage{units}
\hypersetup{colorlinks=true, linkcolor=blue, citecolor=blue, urlcolor=blue}
\newcommand\earth{\hbox{$\oplus$}}
\begin{document}
\title{Direct imaging of a massive dust cloud around R CrB}

%\subtitle{techniques and benchmark results}

\author{
	S.~V.~Jeffers\inst{1}
	\and
	M.~Min\inst{1}
	\and
	L.~B.~F.~M.~Waters\inst{2,3}
	\and
	H.~Canovas\inst{1}
	\and
	M.~Rodenhuis\inst{1}
	\and
	M.~deJuan~Ovelar\inst{1}
	\and
	A.L.~Chies-Santos\inst{1}
	\and
	C.~U.~Keller\inst{1} 
	\thanks{Based on observations made with the William Herschel Telescope operated on the island of La Palma by the Isaac Newton Group in the Spanish Observatorio del Roque de los Muchachos of the Instituto de Astrofísica de Canarias.}}

\offprints{S.~V. Jeffers, \email{S.V.Jeffers@uu.nl}}

\institute{
Sterrekundig Instituut Utrecht,  Utrecht University, P.O.Box 80000, 3508 TA, Utrecht, The Netherlands
\and
SRON, Netherlands Institute for Space Research, 3584 CA Utrecht, The Netherlands
\and
Astronomical Institute "Anton Pannekoek",University of Amsterdam, PO Box 94249, 1090 GE Amsterdam, The Netherlands
}

   \date{\today}

% \abstract{}{}{}{}{} 
% 5 {} token are mandatory
 
  \abstract
  % context heading (optional), leave it empty if necessary  
%   {}
  % aims heading (mandatory)
%   {}
  % methods heading (mandatory)
%   {}
  % results heading (mandatory)
%   {}
  % conclusions heading (optional), leave it empty if necessary 
%   {}
{We present recent polarimetric images of the highly variable star R
CrB using ExPo and archival WFPC2 images from the HST.  We observed R
CrB during its current dramatic minimum where it decreased more than 9
magnitudes due to the formation of an obscuring dust cloud.  Since the
dust cloud is only in the line-of-sight, it mimics a coronograph
allowing the imaging of the star's circumstellar environment.  Our
polarimetric observations surprisingly show another scattering dust
cloud at approximately 1.3" or 2000 AU from the star.  We find that to
obtain a decrease in the stellar light of ~9 magnitudes and with 30\%
of the light being reemitted at infrared wavelengths (from R CrB's
SED) the grains in R CrB's circumstellar environment must have a very
low albedo of approximately 0.07\%.  We show that the properties of
the dust clouds formed around R CrB are best fitted using a
combination of two distinct populations of grains size. The first are
the extremely small 5nm grains, formed in the low density continuous
wind, and the second population of large grains ($\sim$0.14$\mu$m) which
are found in the ejected dust clouds.  The observed scattering cloud,
not only contains such large grains, but is exceptionally massive
compared to the average cloud.}

   \maketitle
%________________________________________________________________

\section{Introduction}

R Coronae Borealis (R CrB) stars are one of the most enigmatic classes
of variable stars.  They frequently show irregular declines in visual
brightness of up to 9 magnitudes due to the production of thick dust
clouds.  Such minima typically last for several months with the
exception of R CrB's current minimum which has lasted for more than 3.5
years (since July 2007 - AAVSO).  The photometric minima, caused by the
obscuration of starlight by a dust cloud, are characterised by a
rapid decline in brightness and a gradual return to normal brightness
levels.  Colour variations are also present with a reddening of the
star during the decline, followed by the star becoming bluer during
the phase of minimum brightness and then becoming redder during the
rise to normal brightness levels.  Variations in colour during a
minimum have also been observed \citep[e.g.][]{efimov88} and are
attributed to pulsations, which are also are present in R CrB's light
curve during maximum light.

%- IR excess
% Clayton 1996 - L band obs don't decrease during obscuration

The colour variations during obscuration can be attributed to the
presence of a cool dust shell, which has been confirmed by observations
of a strong IR-excess \citep{kilkenny84,walker86}.  The cool dust
shell has a temperature of $\approx$ 500--1000 K and L-band infrared
photometric observations show that it still emits in the infrared even
when the line-of-sight is obscured by a dust cloud
\citep{feast86,yudin02}. \cite{feast79} notes that the L-band
observations mirror the pulsation frequencies of the photosphere.  The
cool dust shell is assumed to be continuously replenished by randomly
emitted dust clouds.  The extinction of the ejected dust is different
to that of the ISM, with an UV absorption peak at 2500 \AA\ as opposed to
2175 \AA.  Laboratory measurements and theoretical models show that
this is due to small glassy or amorphous carbon grains (i.e. soot)
formed in a hydrogen-poor environment \citep{hecht84}.

%- NACO images of dust cloud formation

The formation of dust clouds on the R CrB star RY Sgr has been
observed by \cite{laverny04} with NACO at the VLT.  These observations
show that the dust clouds are bright, contributing up to 2\% of the
stellar flux in the infrared (L-band 4.05 $\mu$m), and are located in
any direction at several hundred stellar radii from RY Sgr.
\cite{laverny04} also note that the dust clouds might be very dense
and optically thick close to the star.  \cite{leao07} used
mid-infrared interferometric observation of RY Sgr, with VLTI/MIDI, to
show that the central star is surrounded by a circumstellar envelope
with one bright dust cloud at 100 stellar radii (30 AU).  This is
notably the closest observed dust cloud around an R CrB star.  More
recently \cite{bright11} also used VLTI/MIDI to probe the
circumstellar environments of RY Sgy, V CrA and V854 Cen at very small
spatial scales (50 mas / 400 R$_\star$).  They find that their
observations are consistent with a scenario of random dust ejection
around the star, which over time creates a halo of dust.
  
%\noindent {\bf Dust formation mechanisms:} 

The dust formation mechanism in these stars is not well understood.
However, \cite{crause07} show that there is a correlation between the
onset of brightness declines and R CrB's pulsation period, linking the
expulsion of dust clouds and mass loss to internal stellar processes.
This has also been shown to occur on other R CrB-type stars such as V854
Cen \citep{lawson92} and RY Sgr \citep{pugach77}.  Observations show
broad emission lines (Na D lines, Ca I H \& K and 388.8 nm He line),
as the star's brightness decreases, indicating gas velocities of
$\approx200~$km\,s$^{-1}$.  This is in agreement with observations of
the He I $\lambda$10830 line, which indicate a wind with a velocity of
$200 - 240\,$km\,s$^{-1}$ \citep{clayton03}.  They also note that the P
Cygni profile is present during both maximum light and brightness
decline.  For R CrB, the dust grains are assumed to be formed away
from the star, at approximately 20\,R$_*$, because at the stellar
surface the temperature greatly exceeds the grain condensation
temperature for carbon \citep{feast86,fadeyev88}. The dust clouds are
then driven away from the star by radiation pressure.  An alternative
model proposes that the dust is formed close to the star and moves
quickly away due to radiation pressure \citep[originally proposed
by][]{payne63}.  However, the conditions close to the star are far from
the thermodynamic equilibrium necessary to form dust grains, though as
discussed by \cite{clayton96} and \cite{donn67}, the condensation
temperature of carbon in a hydrogen deficient environment, such as on
an R CrB star, is much higher than when hydrogen is present.

%(though there are two different ideas of thought on
%this).Clayton, G 1992 APJ 397 Empirical arguments for dust formation
%near the stellar surface

To further understand the circumstellar environment of R CrB we have
used imaging polarimetry observations with ExPo, the EXtreme
POlarimeter (Rodenhuis 2012 in preparation), to detect scattered
starlight during its current minimum.  This method is advantageous as
starlight becomes polarised when it is scattered from circumstellar
dust and particularly during obscuration can enable a detection of
close-in circumstellar dust.  In this paper we combine images of the
dust cloud in scattered light, using imaging polarimetry, with
archival images taken by the Hubble Space Telescope (HST) to determine
the properties of the grains in the obscuring and scattering dust
clouds.

In this paper we first describe the observations of R CrB using ExPo
and the HST in Sect. 2 and 3.  In Sect. 4, we summarise the
observational facts that we derive from the observations, which we use
in Sect. 5 to model the properties of the dust grains.  Our results
are discussed in Sect. 6.

\section[]{Polarimetric Imaging with ExPo}

\begin{figure*} 
\begin{center}
\includegraphics[scale=0.55]{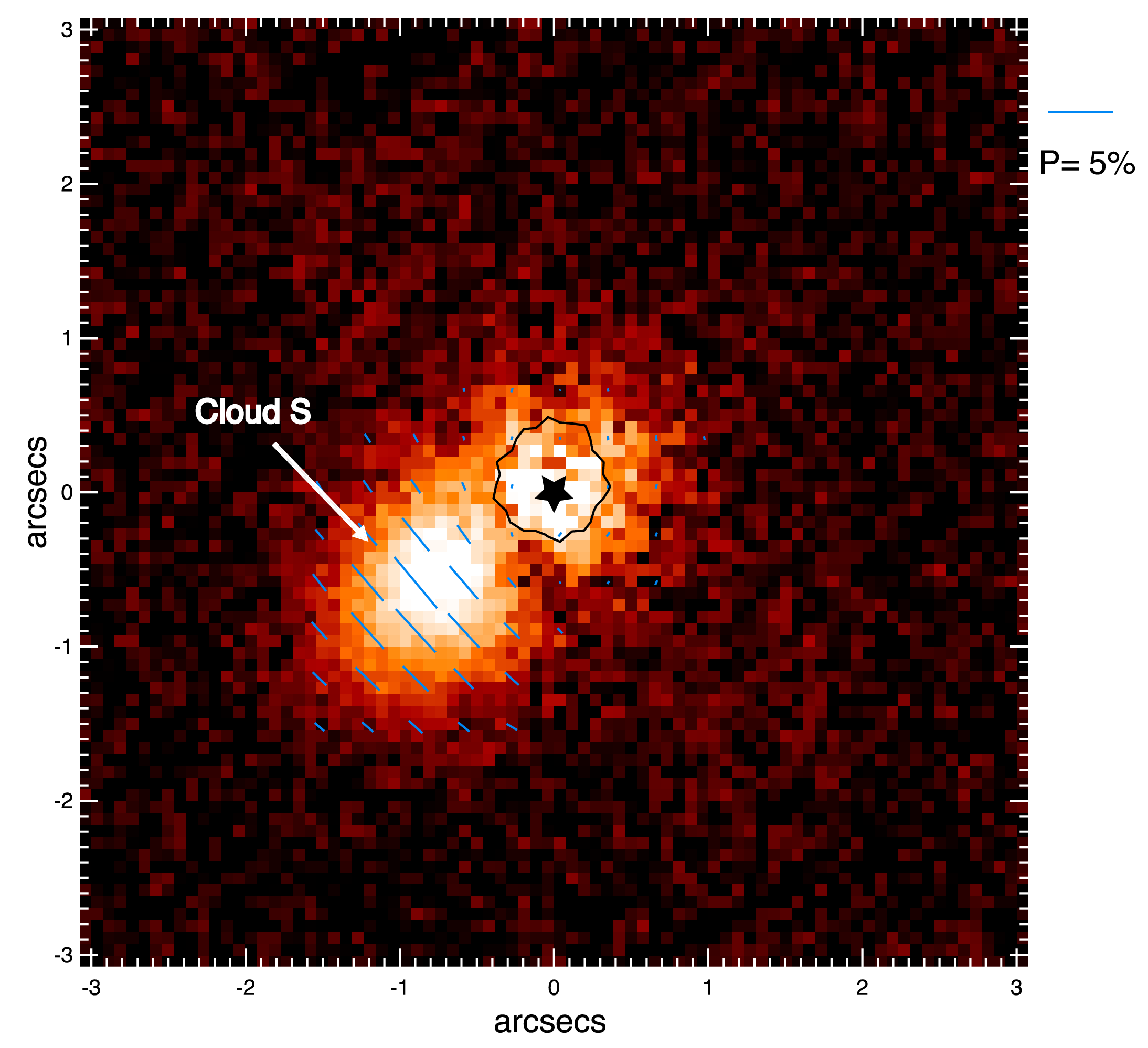}
\end{center}
\caption{ExPo image of the circumstellar environment of R CrB in linear
polarisation.  This image was taken at visible wavelengths (500nm -- 900nm). The scattering dust cloud (Cloud S) is located to the south-east (left) of the image at 1.3" from
the star.  North is upwards and the scale of the image is in arcseconds.}  
\protect\label{expoim} 
\end{figure*}

\subsection{ExPo: Instrumental setup}

The design of ExPo is based on the principle that light
reflected by circumstellar material becomes linearly polarised and can
be easily separated from the unpolarised light that originates from
the central star.  One of the design concepts of ExPo is the
combination of fast modulation of polarisation states with dual-beam
imaging.  This setup significantly reduces the impact of flat field
and seeing effects on the polarimetric sensitivity.

ExPo is a regular visitor instrument at the 4.2m William Herschel
Telescope.  It has been designed and built at Utrecht University and
has a 20"$\times$20" field-of-view and a wavelength range of 500nm to
900nm.  The fast modulation of polarisation states is achieved with
the combination of a Ferroelectric Liquid Crystal (FLC), a cube
beamsplitter and a EM-CCD camera.  The FLC modulates the incoming
light into two polarisation states separated by 90$^\circ$, "A" and
"B" states.  The beam splitter separates each of these frames into two
beams which are imaged on the EM-CCD as $A_\mathrm{left}$ and $A_\mathrm{right}$
followed by $B_\mathrm{left}$ and $B_\mathrm{right}$.

\begin{table}
\caption{Journal summarising ExPo and HST observations}
\protect\label{tab:obs}
\begin{tabular}{l c c c c}
\hline
\hline
Instrument & Date & Filter & Exp time(s) & No exp \\
\hline
ExPo & 27 May 2010 & none & 0.028 & 4x1000 \\ 
% &  & OG590 & 0.028 & 4x1000 \\ 
% &  & Sloan R & 0.028 & 4x1000 \\ 
% &  & Sloan I & 0.028 & 4x1000 \\ \\
HST/WFPC2 & 14 April 2009 & F555W nm & 7 & 1\\ 
    &  & F555W nm & 120 & 4\\ 
    &  & F814W nm & 16 & 1\\ 
    &  & F814W nm & 160 & 3\\ 
\hline
\hline
\end{tabular}
\end{table}

\subsection{ExPo: Observations and data analysis}

Observations of R CrB were secured at the 4.2m William Herschel
Telescope on La Palma, as part of a larger observing run, from 22 to 27
May 2010.  To obtain an image we take typically 4000 frames with an
exposure time of 0.028s per FLC angle i.e. 0$^\circ$, 22.5$^\circ$,
45$^\circ$ and 67.5$^\circ$.  These observations are summarised in
Table ~\ref{tab:obs}. A set of flat fields is taken at the beginning
and end of each night and a set of darks is taken at the beginning of
each observation.

All of the images are dark subtracted, and corrected for flat
fielding, bias and cosmic ray effects. Once the polarization images
are free of instrumental effects, they are carefully aligned and
averaged.  Polarised images are obtained after applying a
double-difference approach to the two beams and alternating "A" and
"B" frames, i.e.
\begin{equation}
 P=0.5 (P_\mathrm{A} - P_\mathrm{B}) = 0.5 ((A_\mathrm{L}-A_\mathrm{R})-(B_\mathrm{L}-B_\mathrm{R}))
\end{equation}

Special care was taken in aligning the R CrB images, since its
polarized image shows one clear dust cloud rather than an extended
polarized source such as a disk or a shell. To minimise guiding
effects, the images were first aligned with a template.  Secondly, to
minimize intensity gradient effects, the images were aligned with an
accuracy of a third of a pixel.  This results in the cloud structure
showing more detail than using other techniques such as aligning the
images with respect to the brightest speckle. Finally, the reduced
images are calibrated using the method of (Rodenhuis et al. 2012 in
preparation) to produce Stokes Q and U images.  The polarised
intensity is defined as $P_\mathrm{I} = \sqrt{Q^2+U^2}$, the degree of
polarization $P = P_\mathrm{I}/I$, where I is the total intensity and the
polarization angle, $P_\Theta = 0.5 arctan \frac{U}{Q}$ defines the
orientation of the polarization plane.  The data analysis is described
in more detail by \cite{canovas11}.  The design of ExPo includes a
polarisation compensator which reduces the instrumental polarisation
of $\sim$3\% to the order of 10$^{-3}$.  This is removed in the data
analysis by assuming that the central star is unpolarised.

The image of R CrB in linear polarisation is shown in
Figure~\ref{expoim}.  To our surprise the image shows one clearly
defined dust cloud,  with a detection of 15$\sigma$, to the southeast
(left) of the star which is indicated by a speckled blob.  The cloud
to the left is the scattering cloud, Cloud S, which shows clearly
defined polarisation vectors.  The systematic / noise error on the
orientation of these vectors is 10$\pm2^\circ$.  The black contour
line at the centre of the image indicates the FWHM of the stellar PSF.

\section{HST Observations}

\begin{figure} \begin{center} 
\includegraphics[scale=0.32]{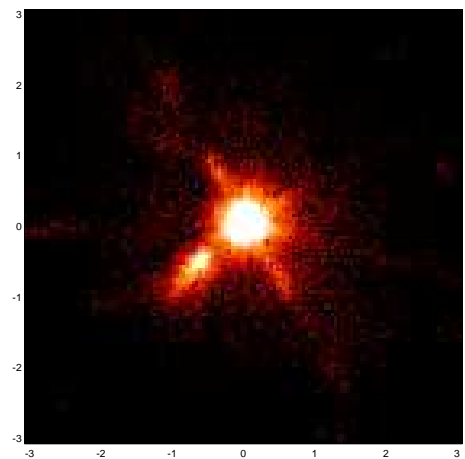} 
\end{center}
\vspace{-0.5cm} \caption{HST/WFPC2 image of the circumstellar
environment of R CrB. This image was taken with the F555W filter
(left).  The scale of the image is
approximately 6x6 arcseconds.}  \protect\label{hstimage1}
\end{figure}

HST images of R CrB were retrieved from the MAST data archive
\citep{clayton11}.  These images were secured on 14 April 2009 using
HST/WFPC2 (Wide Field Planetary Camera) imaging in broad-band V
(F555W) and I (F814W) filters.  These observations are summarised in
Table~\ref{tab:obs}.  The individual exposures taken with each filter
were aligned and combined with the STSDAS task \textit{gcombine} with
a cosmic ray rejection algorithm.  There were 5 images for the V
filter and 4 for the I filter, respectively giving a total exposure
time of 487s and 496s.  Only the Planetary Camera (PC) part of the
images, where R CrB is centred, are used.  The spatial scale for the
PC is $0.046 \arcsec\,pixel^{-1}$ and its field of view is $37\arcsec
\times 37\arcsec$.  The HST observations from April 2009 show an
extended dust cloud (Cloud S) at the same location as shown in the
ExPo images.  The processed images for filters are
shown in Figure~\ref{hstimage1}, magnified to match the field-of-view
of ExPo.  

%\begin{figure*} \begin{center}$ \begin{array}{cc}
%\includegraphics[scale=0.32]{HST555_1.ps} &
%\includegraphics[scale=0.32]{HST814_1.ps} \\ \end{array}$ \end{center}
%\vspace{-0.5cm} \caption{HST/WFPC2 image of the circumstellar
%environment of R CrB. This image was taken with the F555W filter
%(left) and the F814W filter (right).  The scale of the image is
%approximately 6x6 arcseconds.}  \protect\label{hstimage1}
%\end{figure*}

\section[]{Observational Facts}

% summarise the basic facts that we have derived from our ExPo and HST observations

\begin{table}
\caption{Magnitude of R CrB at each epoch of observation and also during its bright unobscured state.}
\protect\label{tab:mag}
\begin{tabular}{l c c c}
\hline
\hline
Filter & 14 April 2009 & 27 May 2010 & Unobscured \\
\hline
V & 15.00 & 14.45 & 5.91 \\
B & 15.65 & 15.15 & 6.49 \\
I & 14.10 & 13.55 & 5.31 \\
\hline
\hline
\end{tabular}
\begin{list}{}{}
\item[{{Note:}}]  Data is from the AAVSO (American Association of Variable Star Observers)
\end{list}
\end{table}

The magnitude of R CrB is shown in Table~\ref{tab:mag} at the epoch of
the ExPo and HST observations.  At the time of observation R CrB is
clearly in a minimum state with $M_V=15.0$, which started in July
2007. The overall reduction in flux, from $M_V=5.91$ to 15.00, is
approximately 4000.  If the obscuring cloud acts like a natural
coronograph completely obscuring the star, then the only contribution
is from light scattered from the dust in the surrounding halo.  To
obtain a magnitude drop of $\approx$ 9 magnitudes the scattered flux
must be less than 1/4000 of the stellar flux. 

\subsection{Photometric variations}

The photometric light curve of R CrB has been monitored for more than
150 years.  During obscuration its light curve is characterised by a
sharp decline followed by a gradual return to maximum light levels.
There are also significant colour changes during obscuration.  Plots
of V versus $B-V$ typically show \citep[e.g. Figure 5 from][]{efimov88} a
very steep decline at the onset of obscuration followed by reddening
and then a shift towards very blue $B-V$ values just before the lowest
light level.  On the return to maximum light levels the light curve
first reddens and then becomes bluer before reaching its normal light
level.  Colour declines that vary from the standard model will be
discussed later in Sect. 5.

%*** FOR INFO *** Plots of the variation of V versus $B-V$ and $B-V$ with
%*** time are shown in the Appendix.

\begin{figure} 
\includegraphics[scale=0.45]{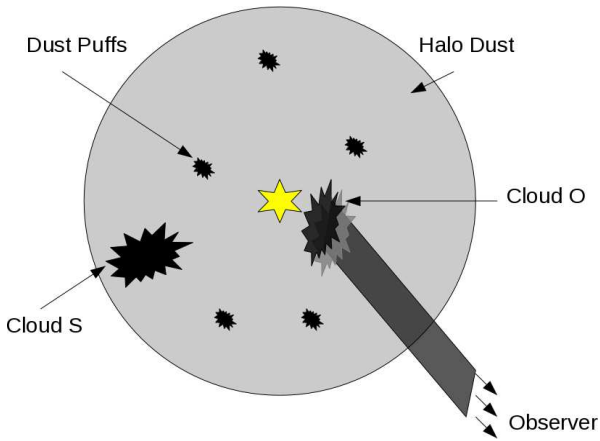}
\caption{Cartoon drawing showing the locations of the three distinct populations of dust around R CrB.}  
\protect\label{dustcartoon} 
\end{figure}

\subsection{Observed dust cloud properties}

The ExPo and HST observations, in scattered light, show that there are
two dust clouds around R CrB.  The first, Cloud O, is the obscuring
cloud which is responsible for the decrease in R CrB's brightness.
The second cloud, Cloud S, the scattering cloud, is detected by HST
and in linearly polarised light by ExPo.  Additionally, in this
analysis we consider that there is a third location of dust in the
halo surrounding the star and is referred to in this work  as Halo
Dust. The locations of these dust populations are shown schematically
in Figure~\ref{dustcartoon}.

\subsubsection{Cloud S}

The properties of Cloud S are determined via aperture photometry (with
a radius=0.47") of the ratio of the flux in Cloud S to the total flux
of the star.  From the ExPo observations, the ratio (total stellar
intensity)/(polarized cloud S intensity)=235. Similarly for the HST
images, the F555W filter gives a Star / Cloud S ratio of 42 and with
the 814nm filter a ratio of 56.

% From the PSF subtracted HST images, this ratio is 3.1 x 10$^{-2}$ in
% the 555nm filter and 1.8 x 10$^{-2}$ in the 814nm filter.  From these
% values the colour and the albedo of the grains can be derived and
% hence the grain sizes.

\subsubsection{Halo dust}

The Halo dust is defined as the halo of dust surrounding R CrB which
is confirmed observationally by L-band observations staying constant
during obscuration.  Previous measurements of the infrared IRAS fluxes
of R CrB (Lambert et al. 2001) show that 30\% of the stellar flux is
reprocessed by absorption of stellar light by the halo grains and
subsequent reemission at infrared wavelengths. In the optically thin
approximation the total amount of reprocessed light in the infrared is
directly proportional to the absorption optical depth through the
equation:
\begin{equation}
f_\mathrm{IR}=1-e^{-\tau_\mathrm{abs}},
\end{equation}

where $\tau_\mathrm{abs}$ is the mean optical depth for absorption
averaged over the stellar spectrum, and $f_\mathrm{IR}$ is the
fraction of the light reprocessed in the infrared. For R CrB,
$f_\mathrm{IR}=0.3$ and consequently $\tau_\mathrm{abs}=0.36$.

The fraction of light that is reprocessed by scattering,
$f_\mathrm{scatt}$, must be extremely small as the integrated
intensity of the system, which is the sum of the starlight and the
scattered light, can drop 9.1 magnitudes in the visual when the star
is obscured. The fraction of scattered light can be equated as
follows:
\begin{equation}
f_\mathrm{scatt}=1-e^{-\tau_\mathrm{scatt}},
\end{equation}

where $\tau_\mathrm{scatt}$ is the scattering optical depth. The
observed decrease is $f_\mathrm{scatt}<1/4000$ (9 magnitudes), which
means that $\tau_\mathrm{scatt}<2.5\cdot10^{-4}$. The single
scattering albedo of the grains must be \begin{equation}
\omega=\frac{\tau_\mathrm{scatt}}{\tau_\mathrm{scatt}+\tau_\mathrm{abs}}<7\cdot10^{-4}.
\end{equation} Consequently, the albedo of the grains in the visual is
extremely small, i.e. on the order of 0.07\%. 

\subsubsection{Cloud O}

The properties of Cloud O are derived from the wavelength-dependent
extinction as the brightness of R CrB decreases.  During R CrB's
bright state the V-band magnitude is 5.91 and its B band magnitude is
6.49, while during obscuration, the V-band and B-band magnitudes are
14.45 and 15.15 respectively (on 27 May 2010).  These values are
summarised in Table~\ref{tab:mag}.

\section[]{Models of Dust Clouds}

%Q-why amorphous carbon?
%Q-how was the grain size distribution varied?

In this Sect. we use the observed properties of the dust clouds to
deduce the properties of the dust grains in Cloud O, Cloud S and in
the Halo Dust surrounding R CrB.  The dust grains are assumed to be
composed of amorphous carbon using the refractive index from
\cite{Preibisch} and the particle shape model from \cite{Min2005}.

\begin{figure}
\begin{center}
\includegraphics[scale=0.35]{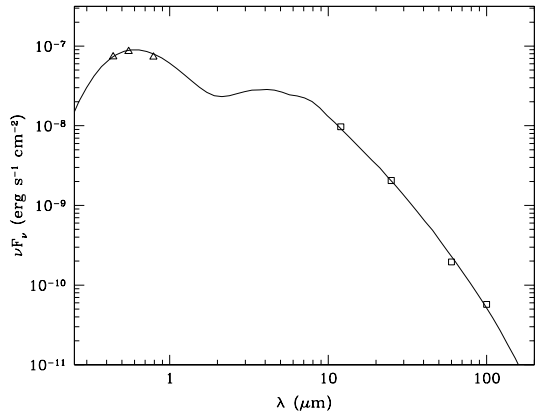} \\ 
\end{center}
\vspace{-0.5cm}
\caption{The modelled fit to the SED of R CrB using the DHS grain model of Min et al.(2005) and the radiative transfer code MCMAX of Min et al. (2009). The optical data points are taken from Table 2.}
\protect\label{f-sed} 
\end{figure}

Cosmic dust grains are not perfect homogeneous spheres. When dust
grains are modelled using Mie theory, i.e. assuming the grains are
perfect homogeneous spheres, the scattering phase function and degree
of polarization show sharp resonances at certain scattering
angles. Furthermore, the overall behavior is very different from that
of more irregularly shaped grains, and even the sign of the
polarization is often wrong \citep[see e.g.][Fig. 10]{munoz04}. Exact
computations of the scattering properties of realistically shaped
grains are computationally demanding and beyond the scope of this
paper. Fortunately, breaking the perfection of a homogeneous sphere in
even the most simple way, suppresses the resonance effects
significantly, for e.g. using the model by \citet{Min2005}. For
the computation of the optical properties we use this model for the
grain shape with $f_\mathrm{max}=0.8$ representing rather irregularly
shaped grains.  This parameter is varied to estimate the error on the
derived parameters.  Throughout the analysis we fix the shape and
composition of the dust grains and vary only their sizes.  Since we
consider irregularly shaped grains we refer to the volume equivalent
radius.

\begin{figure}
\begin{center}$
\begin{array}{cc}
\includegraphics[scale=0.35]{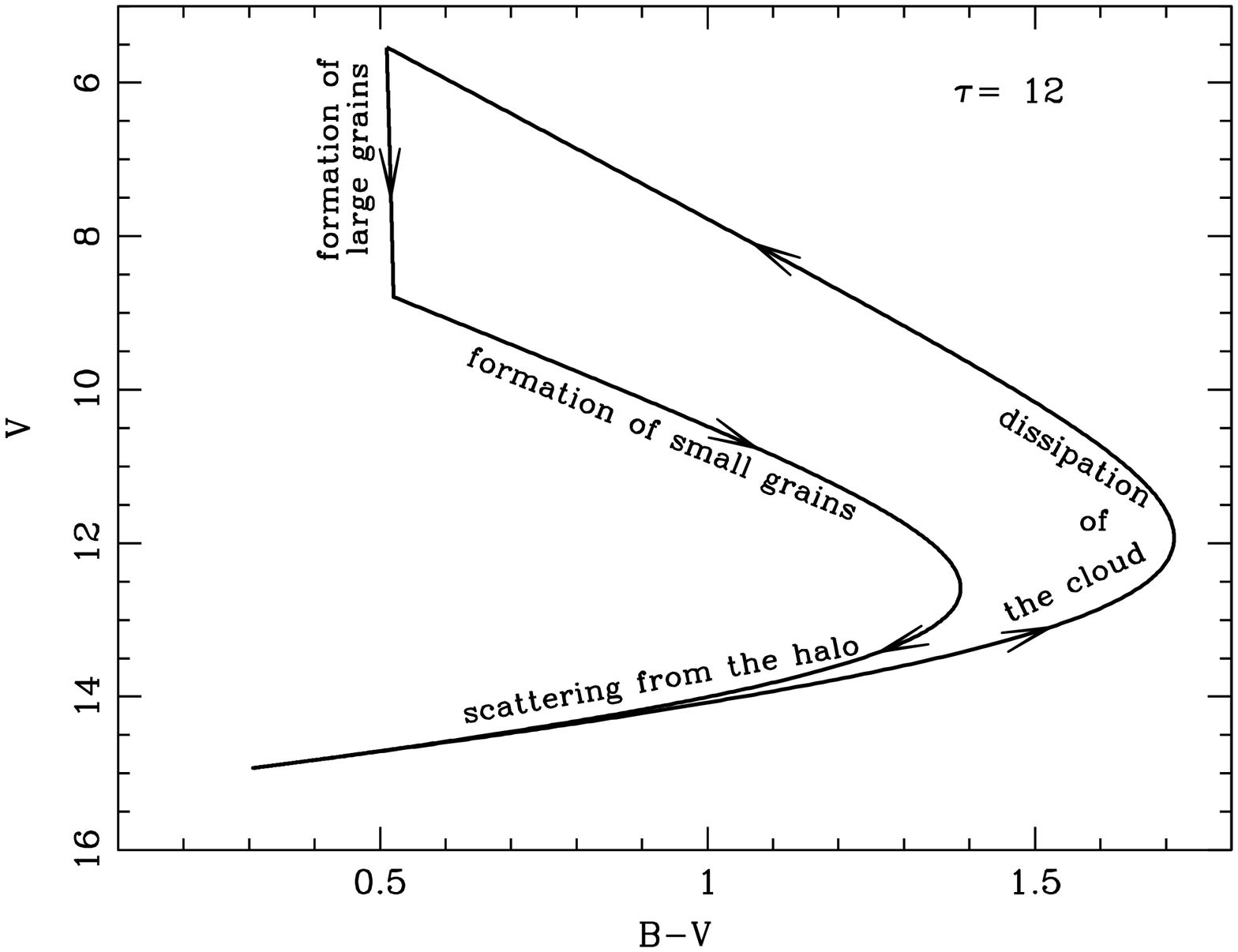} \\
\includegraphics[scale=0.35]{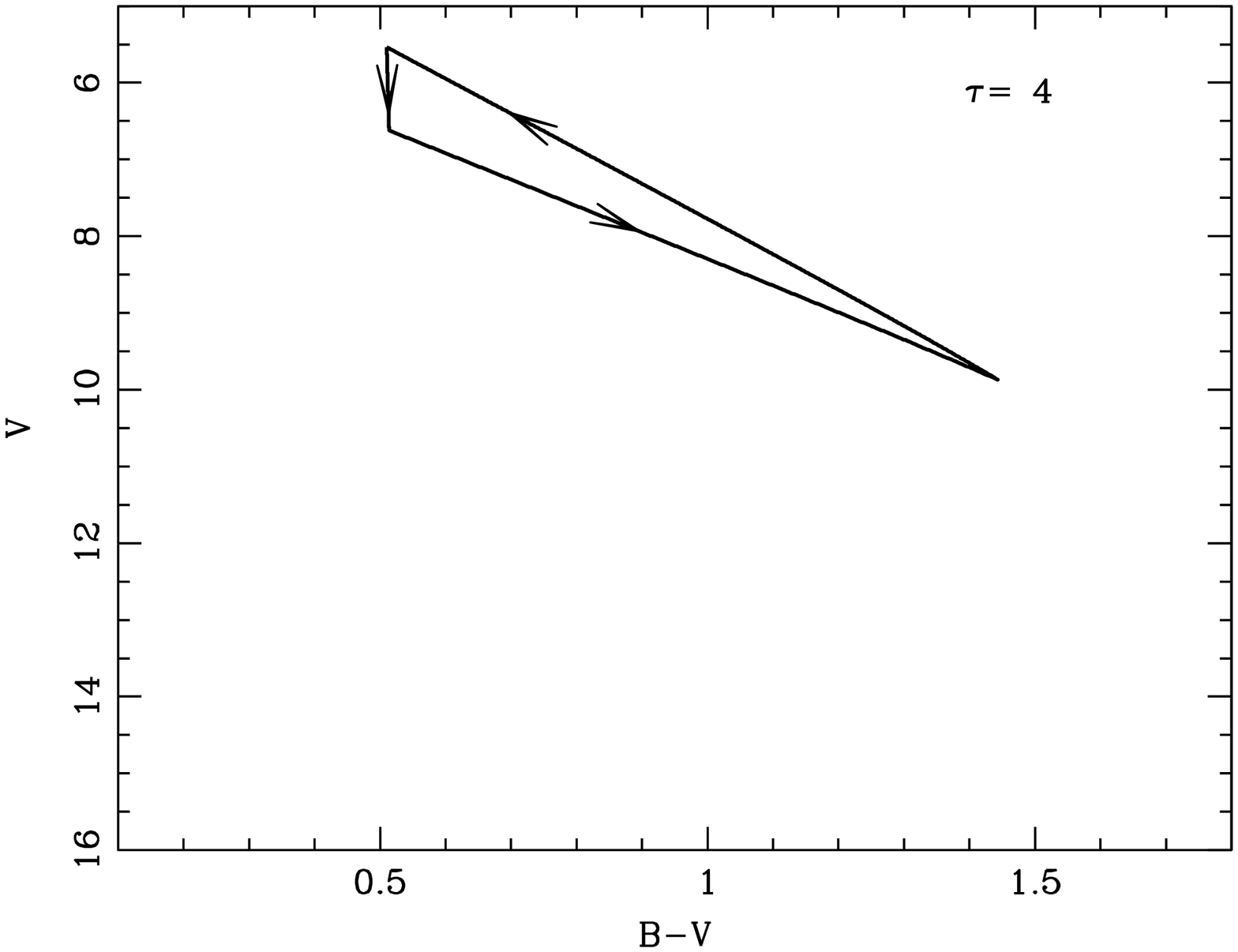} \\
\includegraphics[scale=0.35]{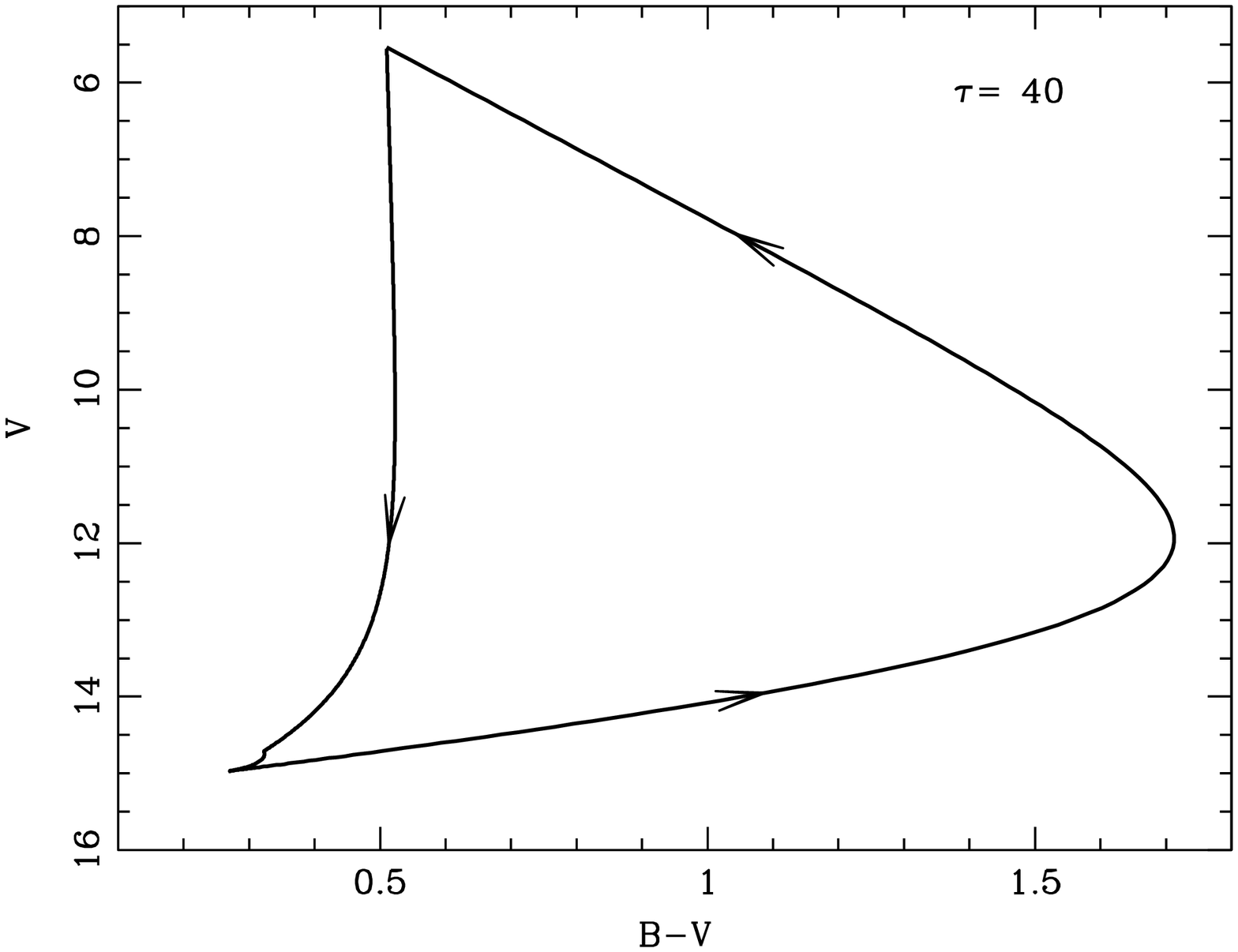}
\end{array}$
\end{center}
\vspace{-0.3cm}
\caption{Models of how the optical depth of the obscuring cloud
impacts the resulting photometric light curve variation during
obscuration, see section 5.2.2 for explanation. For comparison models are shown for $\tau$=12 (normal obscuration),4 (light obscuration), and 40 (dense obscuration) respectively for top, middle and bottom plots.}
\protect\label{f-colvar} 
\end{figure}

\subsection{Halo dust}

To determine the grain properties of Cloud O, it is first necessary to
disentangle the contribution from the surrounding Halo Dust to the
photometric light curve.  As explained in Sect. 4.2.2, the albedo of
a dust grain is very sensitive to the size of the dust grain with
small grains also having a low albedo.  Consequently, it is necessary
to make the grains very small, i.e. smaller than 5 nm to obtain an
albedo of  0.07\%.  This therefore implies that the grains in the
surrounding halo are much smaller than those in Cloud S.

\subsubsection{Fitting the SED}

As previously discussed, to match a dust reprocessing of 30\% and
decrease of 4000 at visible wavelengths it is necessary to assume that
these particles do not scatter very efficiently and consequently are
very small.  To model the SED of R CrB these grains are set to be 5 nm
in radius since R CrB's infrared excess is defined by its Halo dust.
For the star we assume a blackbody with $T_\mathrm{eff}=6750~$K and
$R_* =73.4~R_\odot$, at a distance of 1400 parsec
\citep[from][]{yudin02,asplund97}.  A homogeneous low density dusty
wind is included in the models as a static outflow, with a velocity of
$200~$km\,s$^{-1}$ \citep{clayton03}.  The IRAS fluxes from \cite{walker85} are
fitted by varying the mass loss rate. A dust mass loss rate of
7.5$\times10^{-9} M_\odot yr^{-1}$ is computed to provide the best fit
as shown in Figure ~\ref{f-sed}. This result is similar to that
obtained by \cite{yudin02} though they use a much slower wind (45~km~s$^{-1}$) and consequently determine a lower dust mass loss rate ($3.1\times 10^{-9} M_\odot yr^{-1}$).  The best fit model reproduces the
global properties of the ISO spectrum of \cite{walker99}.

\subsection{Obscuring cloud}
\label{sec:lightcurve}
A characteristic component of the photometric light curve of R CrB is
that there are significant colour changes during obscuration, which
can be used to constrain the grain properties of the obscuring dust
cloud.

\subsubsection{Characteristic light curve properties}

R CrB's light curve typically shows first a sharp decrease in its V
magnitude while $B-V$ remains constant.  This leads to a reddening of
the star as the brightness declines further, followed by the star
becoming bluer until it reaches its minimum.  On the return to normal
brightness levels the light curve first reddens before becoming bluer
\citep[see e.g.][Figure 5 for 1983/4 minimum]{efimov88}.  The
variations in colour are found to vary with the depth of the minimum,
with shallower minima showing the characteristic drop in V, while $B-V$
remains constant, and subsequent red colour variations but are not
followed by the starlight becoming bluer before minimum.  An example
of this is in the shallow minimum of January 1999, where $M_V$
decreases to only 9.5.  For deep obscurations the general behaviour
begins with a sharp decrease in V while $B-V$ remains constant.  It is
only just before the minimum light level that the light curve becomes
bluer before returning to maximum brightness levels following the
standard behaviour i.e. first becoming more red and then more blue.

\begin{figure}
\begin{center}
\includegraphics[scale=0.35, angle=0]{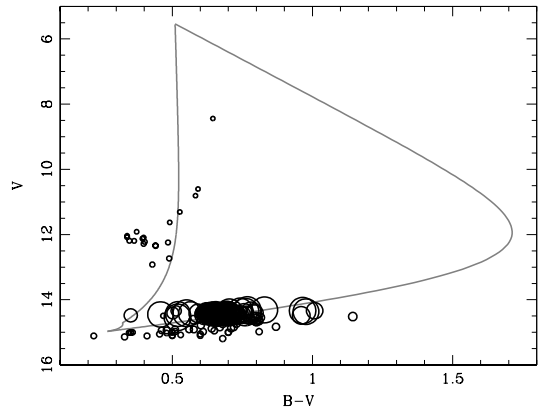} \\ 
\end{center}
\caption{The variation of $B-V$ with decreasing V magnitude of the current 
obscuration, with the $\tau$=40 model overplotted. The size of the plotted 
circle increases with time.  
Note that the first points are not evenly spaced 
in time for the current obscuration. At minimum V magnitude
there are many colour variations of up to $B-V$=0.5 which is an
intrinsic property of R CrB stars.  The photometric data are from the AAVSO.}
\protect\label{f-vbv} 
\end{figure}

\subsubsection{Models of colour variations}

To understand how the sizes of the dust grains in an obscuring `dust
puff' cause the observed colour changes, we have modelled an obscuring
dust cloud from formation to dissipation. The characteristic sharp
drop in V magnitude, while $B-V$ remains constant, indicates the
formation of large grey grains while the dust cloud is still thick
and dense.  This means that the grain sizes are at least $0.2\,\mu$m,
which is the smallest possible size that can give grey extinction.  As
the cloud expands, and consequently has a lower density, only small
particles (i.e. 5 nm) can form, resulting in the onset of the
reddening in the V versus $B-V$ plot.  When the star reaches minimum
light, the cloud is then assumed to dissipate before R CrB returns to
maximum light levels.  The slope of the curve on return to maximum
brightness is determined by the size distribution of the grains.
                                                                        
The obscuring dust cloud is modelled with a total optical depth equal
to 1, 3 and 10 which is only caused by the large grains. The small
grains are added with an optical depth ratio of 3 small grains for
every large grain.  This small:large ratio of 3:1 was chosen to
roughly reproduce the light curve by \cite{efimov88}. Consequently the
total optical depths at minimum light are $\tau=12, 4$ and $40$ the
results of which are respectively shown in Figure~\ref{f-colvar} top,
middle and lower panels.  For the model that uses $\tau$=4, the
minimum brightness level only reaches $M_V=9.9$ and is comparable to
the obscuration in January 1999 where $M_V$ only decreased to 9.5.  As
with all models, large particles are formed at the beginning, with a
sharp vertical decrease in V versus $B-V$, and then small particles are
formed leading to a reddening of the curve.  In this case for low
optical depth, there is no impact from the surrounding Halo dust
cloud.  For $\tau$=12, the curve begins with the same characteristic
behaviour but shows an onset of reddening by small grains beginning at
$\approx$ $M_V=9$, followed by a bluing due to scattering from the
Halo dust at $\approx$ $M_V=12.5$.  The V magnitude then reaches a
minimum of $M_V=14$ before reddening significantly on its return to
maximum brightness levels.  For the model with the highest optical
depth i.e. $\tau$=40, the minimum light can be reached by using only
very large grains (Figure ~\ref{f-colvar} lowest panel).  As with the
other models discussed here, the return to maximum brightness is
characterised by a reddening. This is caused by the dispersal of the
dust cloud, which at this time would also contains a significant
fraction of very small grains.

\subsubsection{Comparison to observed colour variations}

The V versus $B-V$ light curve of the current decline, is shown in
Figure~\ref{f-vbv}. At minimum V magnitude there are many colour
variations of up to $B-V$=0.5 which is an intrinsic property of R CrB
stars (see Sect. 1).  Although the light curve is missing several
data points at the start of the obscuration, the models of the colour
variations are in general agreement with the observed colour
variations.  The best fitting model is for $\tau$=40 which is also
plotted in Figure~\ref{f-vbv}.  To investigate the impact of multiple
scattering we computed a full radiative transfer model for several
points on the V versus $B-V$ plot using MCMAX \citep{min2009}.  The
results show that the inclusion of multiple scattering has the effect
of making the initial sharp drop in V bluer.  To fit the observations,
it is necessary to compensate for this by assuming a smaller grain
size of approximately $0.13\,\mu$m.

\subsection{Scattering cloud}

Observations in scattered light of Cloud S were taken by both HST
and ExPo. In general, the colour of scattering is highly dependent on
the scattering angle, as is the degree of polarization. The light
scattering properties of the dust in Cloud S is determined using the
models of \cite{Min2005}.  To fit the observations it is necessary to
model a grain size and a scattering angle that simultaneously match the polarized intensity of the ExPo images and the colour of the HST
images.

The degree of polarisation is derived to be $>15$\% from the estimate
of the polarized intensity of the ExPo image.  As the ExPo image is
not photometrically calibrated, we estimate the polarized intensity
from the ratio between the total intensity of the star from the
HST image and the polarized intensity from the cloud.  In order to
derive the properties of the cloud we computed the scattering
properties of particles with sizes ranging from $5$\,nm up to
$0.2\,\mu$m with irregularity parameters $f_\mathrm{max}=0.2-0.8$.
The irregularity parameter was varied to determine how it impacts the
derived particle size.  Assuming a single size for the dust grains we
infer that the grains are $\sim$0.14$\pm$0.02\,$\mu$m in diameter. The
error on this value is derived from varying the irregularity parameter
and shows that it has little impact on the derived particle size.  A
significantly smaller grain size results in a colour for Cloud S that
is much bluer than the HST images, while using a larger grain size
produces a result that is too grey. The scattering angle that matches
all of the observations is about $\theta=94\pm33^\circ$ (i.e. Cloud S
is located approximately in the observation plane of the star).  From
our models we can also derive the mass of the cloud which depends on
the scattering efficiency (i.e. particle grain size and the scattering
angle).  We derive a dust mass of the cloud of
$3_{-1}^{+4}\cdot10^{-4} M_{\earth}$ and an intrinsic degree of
polarization of the cloud of $\sim 20\pm5$\%.

\subsection{Summary of models}

It is evident from the SED that there must be a significant dust mass
surrounding R CrB since roughly 30\% of the light is reprocessed as
thermal emission. The very low scattering efficiency of these grains,
as inferred from R CrB's minimum brightness, requires that these
grains are very small.  Contrary to this we find, from the scattered
light by the dust cloud seen in the HST and ExPo images, that the
grains in the dust clouds must be much larger with a size of
approximately $\sim$0.14$\pm$0.02\,$\mu$m . Also, the grey onset of the
obscuration is evidence for the formation of very large grains when
the dust clouds are dense. 
% We find that the grain size of Cloud O is approximately 0.13$\mu$m.

\section{Discussion}

\subsection{Grain sizes}

\subsubsection{Halo dust}

The V-band magnitude of R CrB was 14.45 and 15 respectively at the
epochs of the ExPo and HST observations.  Since R CrB is obscured by a
dust cloud, this magnitude is assumed to be the brightness of the
surrounding Halo dust.  To simultaneously fit the visible magnitude
decrease and the SED, it was therefore necessary to use 5nm grain
sizes since their albedo is very low and are hence very inefficient at
scattering light.  A larger grain size would have a larger albedo and
it would not be possible to obtain the observed decrease in visual
brightness.  The 5nm grain size is at the lower end of the MRN,
Mathis, Rumpl, \& Nordsieck \citep{MathisRumplNordsieck1977}, size
distribution.  An extremely small particle size for the Halo dust of R
CrB has also been found by \cite{yudin02}.  This result is consistent
with the observations of V 854 Cen of \cite{whitney92} where they
state that the scattered flux may arise in the same clouds
contributing to the observed IR flux if the albedo is low.

However, \cite{ohnaka01}, from interferometric observations, show that
the visibility function and spectral energy distribution can be
simultaneously fitted with a model of an optically thin dust shell at
maximum light but not at minimum light.  They suppose that the
discrepancy is attributed to thermal emission from a newly formed dust
cloud, but the results from this paper and that of \cite{yudin02}
conclude that the discrepancy of \cite{ohnaka01} can be resolved by
assuming very small particle sizes, which are very inefficient at
scattering starlight.

\subsubsection{Obscuring cloud}

Currently Cloud O totally obscures the star and we only see the
scattered light from the low density halo making it impossible to
probe its full size distribution. As previously described, the best
fitting model to the observed colour changes is for $\tau$=40, where
total obscuration can be reached with only large grains, and is
overplotted in Figure~\ref{f-vbv}.

\subsubsection{Scattering cloud}

Both the ExPo and HST images of the circumstellar environment of R CrB
clearly show the presence of a large and extended dust cloud, Cloud S.
The cloud is surprisingly elongated and could indicate how dust clouds
interact with their surroundings.  The comprising dust is inferred to
be 0.14$\pm0.2 \mu$m in diameter from combining the light scattering
properties of the dust \citep{Min2005} with the observed colour and
degree of polarisation.  The derived sizes of grains in Cloud O and
Cloud S are in good agreement.  Cloud S is significantly older than
the recently ejected Cloud O, having been ejected at least 50 years
ago,assuming a constant outflow speed of 200 km~s$^{-1}$ (see
Sect. 6.3).  This indicates that there is no significant grain
evolution due to, for example, the high velocity wind that contains
many small particles.  Additionally from the combination of the ExPo
image with the two HST images, we infer that the scattering angle of
the dust is approximately $\theta=94\pm33^\circ$ i.e. almost
perpendicular to the observer.

From our results we derive Halo Dust grain sizes of 5nm, Cloud S grain
sizes of 0.14$\pm 0.2\mu$m and Cloud O grain sizes of 0.13$\mu$m.  To
investigate the impact of using slightly different grain sizes, we
added an size distribution of particles,  modelled by delta
functions at 5 nm and 0.2$\mu$m.  This has the effect of reducing the dust
mass in Cloud S.  However, it is still necessary to include a large
number of these very small grains.  The derived mass of Cloud S
clearly depends on the size of the composing dust grains.  For the
case where the grain size = 0.14$\pm 0.2 \mu$m, the dust cloud mass is
$3_{-1}^{+4}\cdot10^{-4} M_{\earth}$.  This is a minimum mass as
these grains are the most efficient at scattering while still fitting
the colour information from the ExPo and HST images. The mass of Cloud
S when composed of grains that follow the size distribution, can
only be greater than the mass derived for 0.14$\mu$m grains because
this distribution contains many more grains that scatter less
efficiently. However, the increased mass of the model using the
size distribution is perhaps more realistic.

\subsection{Why are there different grain sizes?}

The different grain sizes in Cloud S and O and in the Halo dust could
be explained by models of the formation of dust-driven winds around
late type carbon stars \citep{GailSedlmayr1987} According to this
model, the dust is dominated by very large grains close to the star,
but further out the atoms can only form very small nuclei, due to the
decreased density.  For the case of R CrB, large grains could be
formed for as long as the dust cloud stays sufficiently dense.  If a
significant fraction of the carbon remains after the cloud begins to
disperse, then the remaining carbon will be in a low density
environment and consequently will only be able to form very small
nuclei. This is also what happens in the low density halo, where only
very small grains can form.

\begin{figure}
\begin{center}
\includegraphics[scale=0.35]{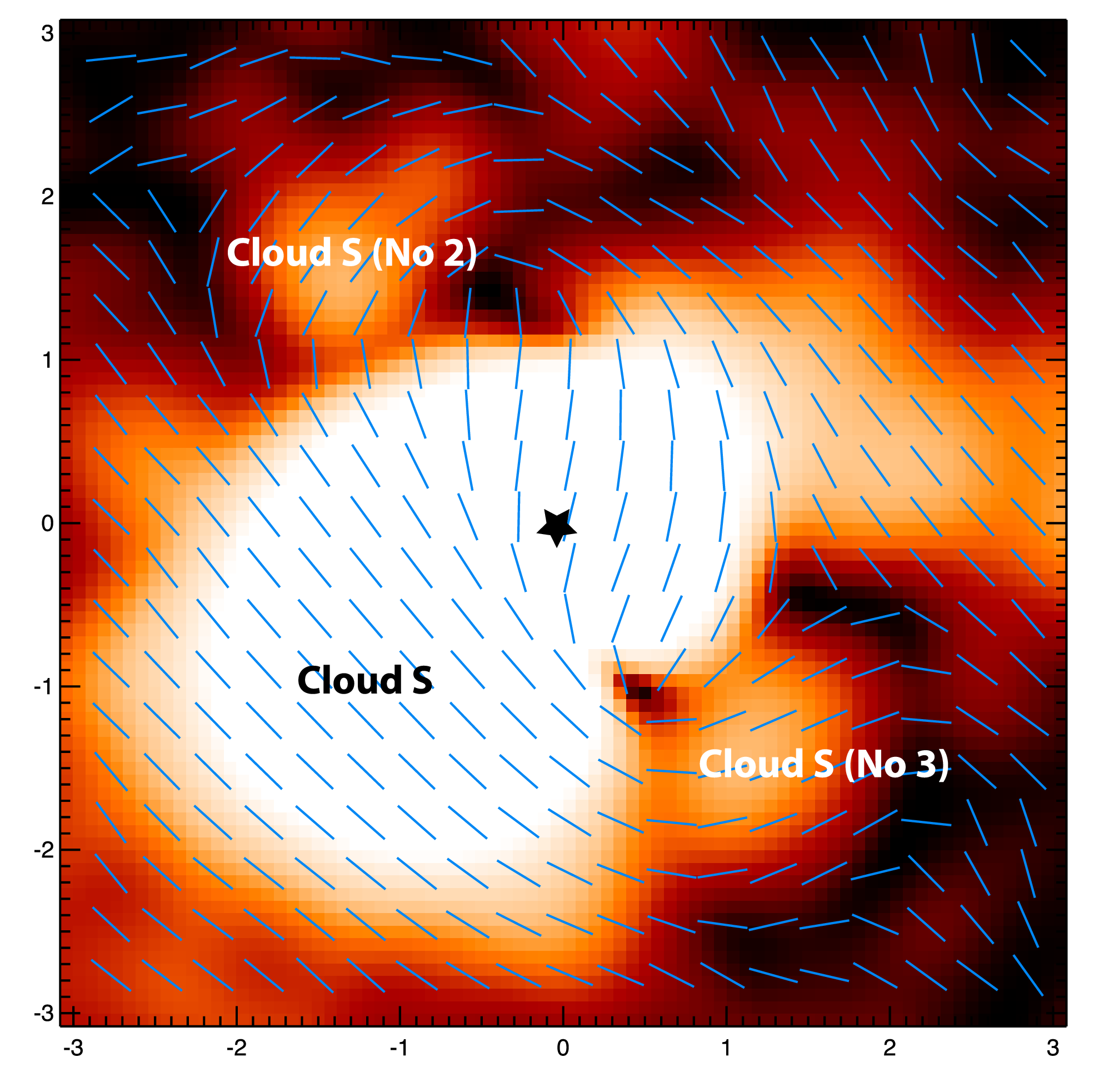} \\
\caption{Smoothed ExPo image to decrease noise effects, showing a tentative detection of additional dust clouds present at 10$^{-5}$ level.}
\label{f-expo2}
\end{center}
\end{figure} 

\subsection{Stellar Wind}

The stellar wind of R CrB has been measured by \cite{clayton03} to be
200 km~s$^{-1}$ from an analysis of the He {\sc i}$\lambda$ 10830 line.
Notably, R CrB shows a P Cygni or asymmetric blue-shifted profile at
all times, i.e. during both minimum and maximum light indicating that
the stellar wind is independent of the ejection of `dust puffs'.  This
is consistent with the model presented in this paper.  The nature of
the stellar wind is highly likely to be dust driven.  According to the
theoretical models of \cite{gail86}, a dust driven stellar wind is
possible in the case of cool Carbon stars with a mass loss rate of the
order of 10$^{-6} M_\odot yr^{-1}$ and with a non-negligible dust
production rate.  The computed dust mass loss rate for R CrB is $0.9\times
10^{-6} M_\odot yr^{-1}$ (from Sect. 5.2.1), though the
conditions for a dust driven wind on R CrB are enhanced since it is
hydrogen deficient star and because there is enormous radiation
pressure on the dust grains.

Polarimetric observations could indicate the presence of permanent
clumpy non-spherical dust shells \citep{clayton97,yudin02}.  Indeed if
the `dust puffs' are ejected from R CrB at a frequency equal to R CrB's
50 day pulsation period \citep{crause07}, at a distance of 2 R$_*$,
there should be a very large number of them between the edge of the
Halo dust and the star.  This frequent ejection of dust puffs in
combination with the dusty stellar wind are considered to be the dust
feeding mechanisms for the circumstellar Halo dust.

\subsection{Dust clouds}

The observations in this paper confirm the `dust puff ejection' model
of R CrB first proposed by \cite{loreta34} and \cite{okeefe39}.  The
HST and ExPo images clearly show a dust cloud at a detection
signal-to-noise of 33 sigma.  Observations of another R CrB star RY
Sgr \citep{laverny04,bright11} show many dust clouds likely to be
randomly ejected from the stellar surface.  As noted by
\cite{laverny04} over the last 50 years the number of brightness
declines for R CrB is much greater than that of RY Sgr, implying that
we should be seeing many more dust clouds in the circumstellar
environment of R CrB.

Tests in the laboratory have shown that ExPo can reach contrast ratios
of up to 10$^{-5}$ which is much fainter than the Cloud S:star ratio
of 10$^{-2}$.  To determine if there are fainter structures also
present in the ExPo images we have smoothed the data using a Gaussian
kernel filter, with a width of 1", to reduce the contribution of
noise.  The resulting image is shown in Figure~\ref{f-expo2}, where
the shown vectors are scaled to the degree of polarisation.  In
addition to the Cloud S, clearly seen in the ExPo images shown in
Figure~\ref{expoim}, there is a tentative detection of two coherent
structures which could be two additional dust clouds.  These clouds
are located just left of centre at the top of the image and just right
of centre at the bottom of the image.  The tentative detection is
plausible because the dust cloud/star contrast ratio is 10$^{-5}$,
they have been detected in all of the ExPo observations and the
polarisation vectors are correctly aligned.  They are unlikely to be
due to spurious instrumental artifacts and do not appear in ExPo
images of stars without circumstellar matter. Future instrumentation
such as SPHERE at the VLT and HiCIAO on Subaru will be able to confirm
this result.

By assuming a `dust puff' formation radius of 2 R$_*$ and a velocity
of 200~km~s$^{-1}$, we derive that Cloud S was ejected 50 years ago.
A surprising factor is that it still remains intact and easily
detectable meaning that it must have been related to an exceptionally
large ejection of dust.  If the cloud has a higher density, due to the
abnormal mass of the cloud, it could form a much bigger fraction of
large, and consequently high-albedo, grains.  If the same age
calculation is applied to the current obscuring cloud, Cloud O, and
assuming that it will continue to travel along the line of sight, it
could imply that R CrB will remain obscured for many decades.

\section{Conclusions}

We conclude that there are two distinct grain populations in the
circumstellar environment of R CrB.  The first is a population of
small 5nm grains that comprise the low density stellar wind and the
second population of large ($\sim$0.14 $\mu$m) grains that are formed
in the ejected dust clouds.  Our polarimetric images together with
archival HST images, surprisingly reveal one exceptionally massive
dust cloud composed of large grains.  The current minimum of R CrB is
also noteworthy and lends additional support to the presence of large
dust grains in ejected dust clouds.

%- do we see other blobs in the ExPo images?  Michiel's image (or are
%they patches of instrumental polarisation) - unlikely - just show one
%image in the discussion

\begin{acknowledgements}
We acknowledge support from NWO.  We also thank the American
Association of Variable Star Observers for the photometric data on R
CrB and the very helpful and knowledgeable staff at the William
Herschel Telescope.
\end{acknowledgements}

\bibliographystyle{aa}
\bibliography{iau_journals,rcrb}

\end{document}